\documentclass[12pt]{article}

\usepackage{graphicx}
\usepackage{amsmath,amssymb}

\begin{document}
\begin{titlepage}
\vfill
\begin{flushright}
{\tt\normalsize KIAS-P08084}\\
{\tt\normalsize SU-ITP-08/35}\\
\end{flushright}
\vfill
\begin{center}
{\Large\bf Holographic Baryons}\footnote{This note is an expanded
version of a proceeding contribution to {\it ``30 years of mathematical
method in high energy physics,"} Kyoto 2008. It will appear in
{\it ``Multifaceted Skyrmion,"  edited by J. Brown and M. Rho, World Scientific.} }

\vfill

Piljin Yi

\vskip 5mm
{\it  School of Physics, Korea Institute for Advanced Study, Seoul 130-722, Korea}
\vfill
\end{center}

\begin{abstract}
\noindent We review baryons in the D4-D8 holographic model of low
energy QCD, with the large $N_c$ and the large 't Hooft coupling
limit. The baryon is identified with a bulk soliton of a unit
Pontryagin number, which from the four-dimensional viewpoint
translates to a modified Skyrmion dressed by condensates
of spin one mesons. We explore classical properties and find that
the baryon in the holographic limit is amenable to an effective
field theory description. We also present a simple method to
capture all leading and subleading interactions in the $1/N_c$ and
the derivative expansions. An infinitely predictive model of
baryon-meson interactions is thus derived, although one may trust
results only for low energy processes, given various
approximations in the bulk. We showcase a few comparisons to
experiments, such the leading axial couplings to pions, the
leading vector-like coupling, and a qualitative prediction of the
electromagnetic vector dominance that involves the entire tower of
vector mesons.

\end{abstract}

\vfill
\end{titlepage}

\renewcommand{\thefootnote}{\#\arabic{footnote}}
\setcounter{footnote}{0}

\section{Low Energy QCD and Solitonic Baryons }

QCD is a challenging theory. Its most interesting aspects, namely the
confinement of color and the chiral symmetry breaking, have defied all
analytical approaches. While there are now many data accumulated from
the lattice gauge theory, the methodology falls well short of giving
us insights on how one may understand these phenomena analytically, nor
does it give us a systematic way of obtaining a low energy theory of
QCD below the confinement scale.

A very useful approach in the conventional field theory language
is the chiral perturbation theory \cite{coleman}.
It bypasses the question of how the confinement and the symmetry breaking occur but rather
focuses on the implications. A quark bilinear condenses to break the
chiral symmetry $U(N_F)_L\times U(N_F)_R$ to its diagonal subgroup $U(N_F)$,
where by $N_F^2$ Goldstone bosons appear, which we will refer to as
pions. They are singled out as the lightest
physical particles, and one guesses and constrains an effective
Lagrangian for them. In the massless limit\footnote{
The effect of small bare masses  for quarks can be
incorporated by an explicit symmetry breaking term
\begin{equation}
{\rm tr} \left( MU+U^\dagger M^\dagger\right)
\end{equation}
with a matrix $M$,
which in our holographic approach would be ignored.}
of the bare quarks, the pions are packaged into a unitary matrix as
\begin{equation}
U(x)=e^{2i\pi(x) /f_\pi}\;,
\end{equation}
whose low energy action is written in a derivative expansion as
\begin{equation}\label{Skyrme}
\int dx^4 \;\left({f_\pi^2\over 4}{\rm tr} \left(U^{-1}\partial_\mu
U\right)^2 +{1\over 32 e^2_{Skyr
me}} {\rm tr} \left[ U^{-1}\partial_\mu U,
U^{-1} \partial_\nu U \right]^2+\cdots\right)\:,
\end{equation}
where the ellipsis denotes higher derivative terms as well as
other possible quartic derivative terms.
One can further add other massive mesons whose masses and interaction
strengths are all left as free parameters to fit with data.

Another analytical approach is
the large $N_c$ expansion \cite{'tHooft:1973jz}.
Here, two different couplings
$1/N_c$ and $\lambda=g_{YM}^2N_c$ control the perturbation expansion, one
counting the topology of the Feynman diagram and the other counting loops.
An interesting question  is how this large $N_c$ limit
appears in the chiral Lagrangian approach. Since the pion fields (or
any other meson fields that one can add) are already color-singlets,
$N_c$ would enter only via the numerical coefficients of the various terms
in the Lagrangian. Both terms shown in (\ref{Skyrme}) can arise from
planar diagrams of large $N_c$ expansion, and we expect
\begin{equation}
f_\pi^{2}\sim N_c\sim \frac{1}{e^2_{Skyrme}} \,.
\end{equation}
Note that since $1/f_\pi^2$ and $1/(e_{Skyrme}^2f_\pi^4)$
play the role of squared couplings for canonically normalized pions,
the self-coupling of pions scales as $N_c^{-1/2}$ \cite{Witten-baryon}.
In particular, this shows that baryons are qualitatively different
than mesons in the large $N_c$ chiral perturbation theory. Baryons involves $N_c$
number of quarks, so the mass is expected to grow linearly with $N_c$, or
equivalently grows with the inverse square of pion self-couplings. In
field theories, such a scaling behavior is a hallmark of nonperturbative
solitons.

Indeed, it has been proposed early on that baryons are
topological solitons, namely Skyrmions \cite{skyrme}, whose baryon number is cataloged by the third
homotopy group of $U(N_F)$, $\pi_3(U(N_F))=Z$.
The topological winding is counted by how many times $U(x)$
covers a noncollapsible three-sphere in $U(N_F)$
manifold, as a
function on $R^3$. Given such topological data, one must find a classical
solution that minimizes the energy of the chiral Lagrangian.
An order of magnitude estimate for the size $L_{Skyrmion}$ of a Skyrmion gives
\begin{equation}
L_{Skyrmion}\sim \frac{1}{f_\pi\, e_{Skyrme}}\;,
\end{equation}
which is independent of large $N_c$.

However, let us pose and consider whether this construction
really makes sense. This solitonic picture says that baryons can be
regarded as coherent states of Goldstone bosons of QCD.
Although the latter are special due to the simple and
universal origin and also due to the light mass, they are one
of many varieties of bi-quark mesons. In
particular, there are known and experimentally measured cubic
couplings between pions and heavier spin one mesons, such
$\rho$ mesons. A condensate of pions, as in a Skyrmion, would
shows up as a source term for a $\rho$ meson equation of motion
and $\rho$ must be also have its own coherent state excited.
In turn, this will disturb the conventional Skyrmion
picture and modify it quantitatively. This is a clear signal
that the usual Skyrmion picture of the baryon has to be modified
significantly in the context of full QCD.

Perhaps because of this, and perhaps for other reasons, the
picture of baryon as Skyrmion have produced mixed results when
compared to experimental data. In this note, we will explore how
this problem is partially cured, in a natural and simple manner
without new unknown parameter, and how the resulting baryons look
qualitatively and quantitatively different from that of Skyrmion.
 As we will see, the holographic picture naturally
brings a gauge-principle in the bulk description of the flavor
dynamics in such a way that all spin one mesons as well as pions
would enter the construction of baryons on the equal footing.
The basic concept of baryon as coherent states of mesons would
remain unchanged, however.
It is the purpose of this note to outline this new approach to
baryons and to explore the consequences.

\section{A Holographic QCD }

A holographic QCD is similar to the chiral perturbation theory
in the sense that we deal with exclusively gauge-invariant operators of
the theory. The huge difference is, however, that this new approach tends
to treat all gauge-invariant objects together. Not only the light meson
fields like pions but also heavy vector mesons and baryons appear together,
at least in principle. In other words, a holographic QCD deals with all
color-singlets simultaneously, giving us a lot more predictive power.
Later we will see examples of this more explicitly.

This new approach is motivated by the large $N_c$ limit of
gauge theories \cite{'tHooft:1973jz} and in particular by the
AdS/CFT correspondence \cite{Maldacena:1997re}. One of the more
interesting notion that emerged in this regard over the last three decades is
the concept of {\it the master field}. The idea is that in the large $N$ limits of
matrix theories with a gauge symmetry, the gauge-singlet observables  behaves
semiclassically in the large $N$ limit \cite{Gopakumar}. Probably the most astounding twist
is the emergence of a new spatial direction in such a picture. As we learned
from AdS/CFT,
{\it the master fields} have to be thought of not as four-dimensional
fields but at least five-dimensional, with the additional direction being
labeled by energy scale. We refer to this new direction as the holographic
direction.

The standard AdS/CFT duality gives us a precise equivalence between the
large $N_c$ maximally supersymmetric Yang-Mills theories and the type IIB
string theory or IIB supergravity in AdS$_5\times S^5$. Here, {\it the master
fields} are nothing but closed string fields such as the gravity
multiplet and excited closed string fields. It is also believed that such
a duality extends to other large $N$ field theories such as ordinary QCD
which is neither supersymmetric nor conformal. The question is then how
to find the right dual theory of the large $N_c$ QCD.

One set of ideas for this, dubbed bottom-up \cite{son}, is similar in spirit
to the chiral perturbation theory. One assumes
that an approximate conformal symmetry exists for a wide range of
energy scales and build up a bulk gravity theory coupled to more
bulk fields, as would be dictated by the AdS/CFT rules if
QCD were conformal. The conformal symmetry is subsequently broken
by cutting off the geometry at both the infrared and the ultraviolet
and by introducing boundary conditions. Necessary degrees of freedoms,
namely the master fields, are introduced as needed by construction,
rather than derived, and in this sense the approach is similar to
the conventional chiral perturbation theory.

The other approach is referred to as top-down, and here one tries to
realized the QCD as a low energy limit of some open string theory
on D-branes, from which a holographic model follows as the closed
string theory dual. Arguably, the best model of this kind we know of is the
D4-D8 system, where $U(N_c)$ D4 gauge theory compactified on a thermal circle
provides large $N_c$ Yang-Mills
sector. The $U(N_F)$ gauge theory on D8 brane, on the other hand,
can be thought of bi-quark meson sector
in the adjoint representation of the $U(N_F)$ flavor symmetry. A crucial
aspect of this model, although to be expected from general AdS/CFT
principles, is that the vector-like flavor symmetry is promoted
to a gauge theory in the bulk.
This D4-D8 model was slowly developed over the years, starting with Witten's initial
identification of the dual geometry for D4 branes wrapped on a thermal circle
\cite{Witten:1998zw}, study of glueball mass spectra of pure QCD without  matter
\cite{Csaki:1998qr,Brower:2000rp}, the
introduction of mesons via D8 branes \cite{sakai-sugimoto}, and very recent
study of baryons as solitonic objects \cite{Hong:2007kx,Hata:2007mb,Hong:2007ay}
on D8 branes. In this section, we will review glueballs and mesons in this D4-D8 model.

\subsection{Holographic Pure QCD from D4}

The story starts with a stack of D4 branes which is compactified
on a circle. The circle here is sometimes called ``thermal" in that
one requires anti-periodic boundary condition on all fermions,
just as one would for the Euclidean time circle when studying finite
temperature field theory. The purpose of having a spatial ``thermal"
circle is to give mass to the fermionic superpartners and thus break
supersymmetry. As is well known, the low energy theory on $N$ D$p$ branes
is a maximally supersymmetric $U(N)$ Yang-Mills theory in $p+1$ dimensions,
so putting $N_c$ D4 branes on a thermal circle, we obtains pure $U(N_c)$
Yang-Mills theory in the noncompact $3+1$ dimensions. We are interested in
large $N_c$ limit, so the $U(1)$ part can
be safely ignored, and we may pretend that we are studying $SU(N_c)$
theory instead. While the anti-periodic boundary condition generates
massgap only to fermionic sector at tree level, scalar partners also become
massive since there is no symmetry to prohibit their mass any more. Only
the gauge multiplet is protected.

We then extrapolate the general idea of AdS/CFT to this non-conformal
case, which states that, instead of studying strongly coupled
large $N_c$ Yang-Mills theory, one may look at its dual closed string
theory. The correct closed string background to use is nothing but the string
background generated by the D4 branes in question. This geometry was first
written down by Gibbons and Maeda
\cite{Gibbons:1987ps} in the 1980's, and later reinterpreted by Witten in 1998
as the dual geometry for D4 branes on a thermal circle \cite{Witten:1998zw}.
The metric is most conveniently written as
\begin{equation}
ds^2=\left(\frac{U}{R}\right)^{3/2}\left(\eta_{\mu\nu}dx^{\mu}dx^{\nu}+f(U)d\tau^2\right)
+\left(\frac{R}{U}\right)^{3/2}\left(\frac{dU^2}{f(U)}+U^2d\Omega_4^2\right) \;,
\end{equation}
with $R^3=\pi g_sN_cl_s^3$ and $f(U)=1-U_{KK}^3/U^3$.
The topology of the spacetime is $R^{3+1}\times D\times S^4$,
with the coordinate $\tau$ labeling the azimuthal angle of the disk $D$,
with $\tau=\tau+\delta\tau$ and
$\delta\tau=4\pi R^{3/2}/(3U_{KK}^{1/2})$. The circle parameterized by $\tau$
is the thermal circle.
The dilaton is
\begin{equation}
e^{-\Phi}=\frac{1}{g_s}\left(\frac{R}{U}\right)^{3/4} \:,
\end{equation}
while the antisymmetric Ramond-Ramond background field $C_{3}$
is such that $dC_3$ carries $N_c$ unit
of flux along $S^4$.

\begin{figure}[t]
\begin{center}
\scalebox{0.5}[0.5]{\includegraphics{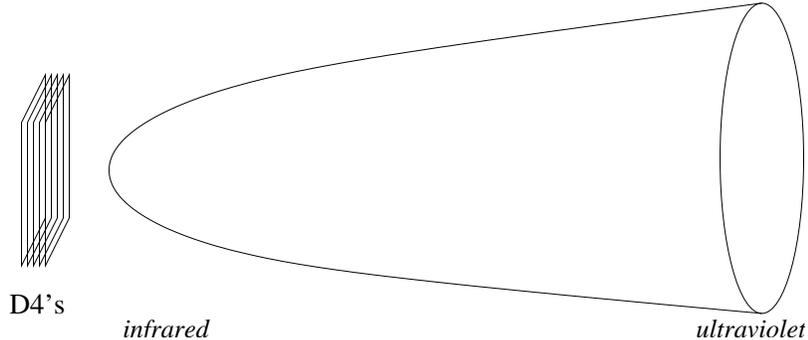}}
\par
\vskip-2.0cm{}
\end{center}
\caption{\small A schematic diagram showing the dual geometry.
A stack of D4's responsible for the dual geometry are shown
for an illustrative purpose, although the actual spacetime
does not include them. The manifold shown explicitly is
spanned by the angle $\tau$ and the radial coordinate $U$.
The thermal circle spanned by $\tau$  closes itself
in the infrared end due to the strong interaction of QCD.
Small excitations of
metric (and its multiplet) at the infrared end correspond to
glueballs. } \label{fig}
\end{figure}

In the limit of large curvature radius, thus large $N_c$, and in the
limit of large 't Hooft coupling $\lambda\equiv g_{YM}^2N_c$, the duality
collapse to a relationship between the theory of D4 branes to type IIA
supergravity defined in this background. Given the lack of useful method of
string theory quantization in curved background, this is the best we can
do at the moment. Therefore, all computations in any of holographic QCD
must assume such a limit and extrapolate to realistic regime at the end
of the computation. This is also the route that we follow in this note.

Among remarkable works in early days of AdS/CFT is the study of glueball
spectra in this background \cite{Csaki:1998qr,Brower:2000rp}.
They considered small
fluctuations of IIA gravity multiplet in the above background, with the
plane-wave like behavior along $x^\mu$ and $L^2$ normalizability along the
remaining six directions. They identified each of such modes as glueballs
up to spin 2, and computed their mass$^2$ eigenvalues as dictated by the
linearized gravitational equation of motion.

This illustrates
 what is going on here. We can think of the duality here as a
simple statement that the open string side and the closed string side is one
and the same theory. The reason we have apparently more complicated description
on the open string side is because there we started with a misleading and
redundant set of elementary fields, namely the gauge field whose number
scales as $N_c^2$, only
to be off-set by the gauge symmetry. The closed
string side, or its gravity limit, happens to be more smart about what are
the right low energy degrees of freedom and encodes only gauge-invariant
ones. For pure Yang-Mills theory like this, the only gauge-invariant
objects are glueballs, so the dual gravitational side should compute
the glueball physics.

The expectation that there exists a more intelligent theory consisting
only of gauge-invariant objects in the large $N_c$ limit
is thus realized via string theory in a somewhat
surprising manner that {\it the master fields}, those truly physical
degrees of freedom, actually live not in four
dimensional Minkowskian world but in five or higher dimensional curved
geometry. This is not however completely unanticipated, and was heralded
in the celebrated work by Eguchi and Kawai in early 1980's \cite{Eguchi:1982nm}
which is all the more remarkable in retrospect.
For the rest of this note, we will continue this path and try to incorporate
massless quarks to the story.

\subsection{Adding Mesons via D4-D8 Complex}

To add mesons, Sakai and Sugimoto introduced the $N_F$ D8 branes, which share
the coordinates $x^\mu$ with the above D4 branes \cite{sakai-sugimoto} and are
transverse to the thermal circle $\tau$.
Before we trade off the $N_c$ D4 branes in favor of the dual gravity
theory, this would have allowed massless quark as open strings ending on
both the D4 and the D8 branes. As the D4's are replaced by the dual geometry,
however, the 4-8 open strings have to be paired up into 8-8 open strings,
which are naturally identified as bi-quark mesons. From the viewpoint of
D8 branes, the lightest of such mesons belong to a $U(N_F)$ gauge field.

The  $U(N_F)$ gauge theory on D8 branes has the action
\begin{equation}
-\frac{4\pi^2l_s^4\mu_8}{8}\int \sqrt{-h_{8+1}}\;e^{-\Phi}\;{\rm tr} {\cal F}^2
+
\mu_8\int\,C_3 \wedge {\rm tr}\,e^{2\pi\alpha' {\cal F} }\,,
\end{equation}
where the contraction is via the induced metric of D8 and
$\mu_p={2\pi}/{(2\pi l_s)^{p+1}} $
with  $l_s^2=\alpha'$.  The induced metric on the D8 brane is
\begin{equation}
h_{8+1}=\frac{U^{3/2}(w)}{R^{3/2}}\left(dw^2+\eta_{\mu\nu}dx^{\mu}dx^{\nu}\right)
+\frac{R^{3/2}}{U^{1/2}(w)} d\Omega_4^2\:,
\end{equation}
after we trade off the holographic (or radial) coordinate $U$ in favor of
a conformal one $w$ as
\begin{equation}
w=\int_{U_{KK}}^U{R^{3/2}dU^\prime}/{\sqrt{{U^\prime}^3-U_{KK}^3}}\:,
\end{equation}
which resides in a finite interval of length $\sim O(1/M_{KK})$ where
$
M_{KK}\equiv 3U_{KK}^{1/2}/2R^{3/2} \:.
$
Thus, the topology of the $D8$ worldvolume is $R^{3+1}\times I\times S^4$.
The nominal Yang-Mills coupling $g_{YM}^2$ is related to the other
parameters as
\begin{equation}
g_{YM}^2=2\pi g_s M_{KK} l_s \;,
\end{equation}
which is not, however, a physical parameter on its own. The low energy
parameters of this holographic theory are $M_{KK}$ and $\lambda$, which
together with $N_c$ sets all the physical scales such as the QCD scale
and the pion decay constant.

\begin{figure}[t]
\begin{center}
\scalebox{0.5}[0.5]{\includegraphics{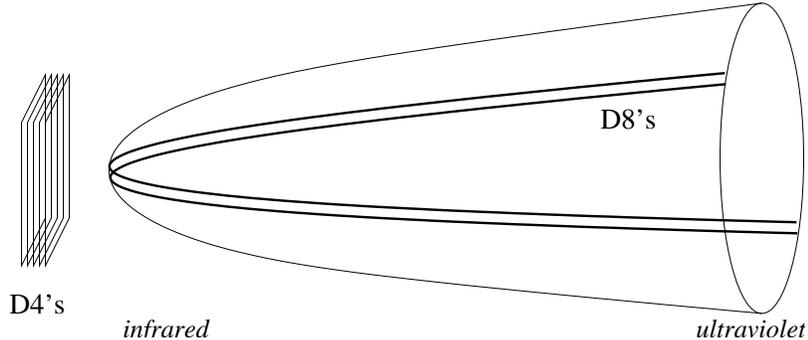}}
\par
\vskip-2.0cm{}
\end{center}
\caption{\small The firgure shows how D8's are added to the system.
Low energy excitations (also located at the infrared end) of D8-D8
open strings are bi-quark mesons. } \label{fig}
\end{figure}

In the low energy limit,  we  ignore $S^4$ direction on
which D8's are completely wrapped, and find a
five-dimensional Yang-Mills theory with a Chern-Simons term
\begin{eqnarray}\label{five}
-\frac14\;\int_{4+1}
\; \frac{1}{e(w)^2}\sqrt{-h_{4+1}}\;{\rm tr} {\cal F}^2
+\frac{N_c}{24\pi^2}\int_{4+1}\omega_{5}({\cal A})\: ,\label{dbi}
\end{eqnarray}
where the position-dependent Yang-Mills coupling of this flavor
gauge theory is
\begin{equation}
\frac{1}{e(w)^2}=\frac{e^{-\Phi}V_{S^4}}{2\pi (2\pi l_s)^5}
=
\frac{\lambda N_c}{108\pi^3}M_{KK}\frac{U(w)}{U_{KK}}\:
\end{equation}
with $V_{S^4}$ the position-dependent volume of $S^4$.
The Chern-Simons coupling with $d\omega_5({\cal A})={\rm tr} {\cal F}^3$
arises because $\int_{S^4}dC_3\sim N_c$.

As advertised, this by itself generates many of bi-quark mesons of
QCD. More  specifically, all of vector and axial-vector mesons
and the pion multiplet are encoded in this five-dimensional $U(N_F)$
gauge field. The vector mesons and the axial vector mesons are more
straightforward conceptually, since any ``compactification" of five-dimensional
Yang-Mills theory would lead to an infinite tower of four-dimensional
massive vector fields.
Although the radial direction $w$ (or $U$) is infinite in terms of
proper length, equation of motion is such that normalizable fields
are strongly pushed away from the boundary, making it effectively a
compact direction. The usual Kaluza-Klein reduction
(in the somewhat illegal but convenient axial gauge
${\cal A}_w=0$),
\begin{equation}\label{ex}
{\cal A}_\mu(x;w)= i\alpha_\mu(x)\psi_0(w)+i\beta_\mu(x) +\sum_n
a_\mu^{(n)}(x)\psi_{(n)}(w)\:
\end{equation}
contains an infinite number of vector fields, whose action can
be derived explicitly as,
\begin{equation}
\int dx^4\,{\cal L}=\int dx^4 \sum_n{\rm tr}\; \left\{{1\over 2} {\cal F}_{\mu\nu}^{(n)}
{\cal F}^{\mu\nu(n)}+m_{(n)}^2 a_\mu^{(n)} a^{\mu(n)}\right\}+\cdots\:,
\end{equation}
with ${\cal F}^{(n)}_{\mu\nu}=\partial_\mu a^{(n)}_\nu-\partial_\nu
a^{(n)}_\mu$. The ellipsis denotes zero mode part, to be discussed shortly,
as well as infinite number of couplings among these infinite varieties of mesons,
all of which come from the unique structure of the
five-dimensional $U(N_F)$ Yang-Mills Lagrangian in (\ref{five}). Because ${\cal A}$
has a specific parity, the parity of $a_n$'s are determined by the
parity of the eigenfunctions $\psi_{(n)}(w)$ along the fifth direction.
Since the parity of any one-dimensional eigenvalue system alternates,
an alternating tower of vector and axial-vector fields emerge as the
masses $m_{(n)}$ of the KK modes increase.

For each such eigenmode, five-dimensional massless vector field has three
degrees of freedom, so is natural for massive four-dimensional vector fields
to appear.
An exception to this naive counting, which is specific to the gauge theory,
is the zero mode sector. In Eq.~(\ref{ex}), we separated it out from the rest
as $\alpha(x)$ and $\beta(x)$ terms. To understand
this part, it is better to give up the axial gauge and consider the Wilson
line,
\begin{equation}
U(x)=e^{i\int_w {\cal A}(x,w)} \;,
\end{equation}
which, as the notation suggests, one identifies with the pion field $
U(x)=e^{2i\pi(x) /f_\pi}$.
Upon taking a singular gauge transformation back
to ${\cal A}_w=0$, one finds that it is related to $\alpha$ and $\beta$ as
\begin{eqnarray}
\alpha_\mu(x)\equiv  \{U^{-1/2},\partial_\mu U^{1/2}\} \;,\quad
2\beta_\mu(x)\equiv [U^{-1/2},\partial_\mu U^{1/2}] \;.
\end{eqnarray}
Truncating to this zero mode sector reproduces a Skyrme
Lagrangian of pions \cite{skyrme} as a dimensional reduction of the
five-dimensional Yang-Mills action,
\begin{equation}\label{Skyrme}
\int dx^4 \;\left({f_\pi^2\over 4}{\rm tr} \left(U^{-1}\partial_\mu
U\right)^2 +{1\over 32 e^2_{Skyr
me}} {\rm tr} \left[ U^{-1}\partial_\mu U,
U^{-1} \partial_\nu U \right]^2\right)\:,
\end{equation}
with $ f_\pi^2=(g_{YM}^2 N_c) N_c M_{KK}^2/54\pi^4$ and
$1/e^2_{Skyrme}\simeq  {61 (g_{YM}^2 N_c)N_c}/54\pi^7$.
No other quartic term arise, nor do we find higher order terms
in derivative, although we do recover
the Wess-Zumino-Witten term
from the Chern-Simons term \cite{sakai-sugimoto}.
To compare against actual QCD, we must fix $\lambda=g_{YM}^2N_c\simeq 17$ and
$M_{KK}\simeq 0.94\, GeV$
to fit  both the pion decay constant $f_\pi$ and the mass of the first vector  meson.
After this fitting, all other infinite number of masses and coupling constants
are fixed. This version of the holographic QCD is extremely predictive.

Let us emphasize that the meson system
here comes with a qualification. Note that we treated D8 branes
differently than D4 branes. The latter are replaced by the dual
geometry while the former are kept as branes. This has to be because
we are interested in objects charged under $U(N_F)$, whereas we
are only interested in singlets under $U(N_c)$. However, we not only
treated D8 as branes but also as probe branes, meaning that
the backreaction of D8 to the dual geometry of D4's is ignored.
In terms of field theory language, we effectively ignored Feynman
diagrams involving quarks in the internal lines, resulting in
the quenched approximation.

\section{Holographic Baryons}

The baryon can be naturally
regarded as a coherent state of mesons in the large $N_c$. In the
conventional chiral Lagrangian approach, is the Skyrmion made from
pions, which we argued cannot be the full picture. In D4-D8 model
of holographic QCD above, especially, pions are only zero mode
part of a holographic flavor theory, and infinite towers of vector
and axial-vector mesons are packaged together with pions into a single
five-dimensional $U(N_F)$ gauge field.
This suggests that the picture of baryon as a soliton must be
lifted to a five-dimensional soliton of this  $U(N_F)$  gauge theory
in the bulk, in such a manner that spin one mesons contribute to
construction of baryons as well. In this section, we explore
classical and quantum properties of this holographic and
new version of Skyrmion.

\subsection{The Instanton Soliton}

The five-dimensional effective action for the $U(N_F)$ gauge field
in Eq.~(\ref{five}) admits solitons which carry a Pontryagin number
\begin{equation}
\frac{1}{8\pi^2}\int_{R^3\times I} {\rm tr} F\wedge F=k \: ,
\end{equation}
with integral $k$. We denoted by $F$ the non-Abelian part of
${\cal F}$ (and similarly later, $A$ for non-Abelian part of ${\cal A}$).
The smallest unit  with $k=1$ turns out
to carry quantum numbers of the baryon. The easiest way to see this identification
is to relate it to the Skyrmion \cite{skyrme} of the chiral perturbation theory .

Recall that both instantons and Skyrmions are labeled by the
third homotopy group $\pi_3$ of a group manifold, which is the
integer for any semi-simple Lie group manifold $G$. For the
Skyrmion, the winding number show up in the classification of maps
\begin{equation}
U(x):R^3\rightarrow SU(N_F) \;.
\end{equation}
For the instanton whose asymptotic
form is required to be pure gauge,
\begin{equation}
A(x,w\rightarrow\pm\infty)=ig_\pm(x)^\dagger dg_\pm(x) \;,
\end{equation}
the winding number is in the classification of the map
\begin{equation}
g_-(x)^\dagger  g_+(x):R^3\rightarrow SU(N_F) \;.
\end{equation}
The relationship between the two types of the
soliton is immediate \cite{Atiyah:1989dq}. Recall that the $U$ field of chiral
perturbation theory is obtained in our holographic picture
as the open and infinite Wilson line along $w$ direction.
On the other hand, the Wilson line computes nothing but
$g_-(x)^\dagger g_+(x)$, so we find that
\begin{equation}
U_k(x)=e^{i\int_w {A}^{(k)}(x,w)}
\end{equation}
carries $k$ Skyrmion number exactly when $A^{(k)}$ carries
$k$ Pontryagin number. Therefore, the instanton soliton
in five dimensions is  the holographic image of the Skyrmions in
four dimensions. We will call it the instanton soliton.

Normal instantons on a conformally flat four-manifold are well
studied, and the counting of zero
modes says that for $k$ instanton in $U(N_F)$ theory, there
are $4kN_F$ collective coordinates. For the minimal case with $k=1$
and $N_F=2$, giving us eight collective coordinates. They are
four translations, one overall size, and three gauge
rotations. For our instanton solitons, this counting does not
hold any more.

Unlike the usual Yang-Mills theory in trivial $R^4$ background,
the effective action has a position-dependent
inverse Yang-Mills coupling $1/e(w)^2$ which is a monotonically
increasing function of $|w|$. Since the Pontryagin density
contributes to action as multiplied by $1/e(w)^2$, this tends to
position the soliton near $w=0$ and also shrink it for the same
reason. The $F^2$ energy of a trial configuration with size $\rho$
can be estimated easily in the small $\rho$ limit,\footnote{ The estimate of energy
here takes into account the spread of the instanton density
$D(x^i,w)\sim \rho^4/(r^2+w^2+\rho^2)^4$, but ignores
the deviation from the flat geometry along the four
spatial directions.}
\begin{equation}
E_{\rm Pontryagin}= \frac{\lambda N_c}{27\pi}M_{KK}\times \left(1+
 \frac16\, M_{KK}^2\rho^2+\cdots\right)\:,
\end{equation}
which clearly shows that the energy from the kinetic term increases
with $\rho$. This by itself would collapse the soliton to a
point-like one, making further analysis impossible.

A second difference comes from the presence of the additional
Chern-Simons term $\sim {\rm tr} {\cal A\wedge F\wedge F}$, whereby
the Pontryagin density $F\wedge F$ sources some of the gauge
field ${\cal A}$ minimally.
This electric charge density costs the Coulombic energy
\begin{equation}\label{C}
E_{\rm Coulomb}\simeq \frac12\times \frac{e(0)^2N_c^2}{10\pi^2\rho^2}+\cdots\:,
\end{equation}
again in the limit of $\rho M_{KK} \ll 1$. This Coulombic energy
tends to favor larger soliton size, which competes against the
shrinking force due to $E_{Pontryagin}$.

The combined energy is minimized at \cite{Hong:2007kx,Hata:2007mb,Hong:2007ay}
\begin{equation}\label{size}
\rho_{baryon}\simeq \frac{({2\cdot 3^7\cdot\pi^2/5})^{1/4}}{M_{KK}\sqrt\lambda }\:,
\end{equation}
and the  classical mass of the stabilized soliton  is
\begin{eqnarray}\label{mass}
m_B^{classical}&=&\left(E_{\rm Pontryagin}+E_{\rm Coulomb}\right)\biggr\vert_{\rm minimum} \nonumber\\
&=&\frac{\lambda N_c}{27\pi}M_{KK}
\times\left(1+\frac{\sqrt{2\cdot 3^5\cdot \pi^2/5}}{\lambda} +\cdots\right)\;.
\end{eqnarray}
As was mentioned above, the size $\rho_{baryon}$ is significantly
smaller than $\sim 1/M_{KK}$. We have a
classical soliton whose size is a lot smaller than the fundamental
scale of the effective theory. On the other hand,
this small soliton size is still much larger than its own Compton size
$1/m_B^{classical}\simeq 27\pi/(M_{KK}\lambda N_c)$, justifying our
assertion that this is indeed a soliton.

Note that the instanton soliton size is much smaller that the
Skyrmion size when  the 't Hooft coupling is
large.\footnote{One must not confuse these solitonic sizes with
the electromagnetic size of baryons. The latter is dictated
by how photons interact with the baryon, and in the holographic
QCD  with $\lambda\gg 1$ is determined at $\rho $ meson scale
and independent of $\lambda$, due to the vector dominance.
One may think of these solitonic sizes as being hadronic.}  We already saw
that the Skyrmion size is determined by the ratio of the two
dimensionful couplings in the chiral Lagrangian. Using the
values of these coupling derived from our D4-D8 model, the
would-be Skyrmion size is
\begin{equation}
L_{Skyrmion}\sim \frac{1}{f_\pi e_{Skyrme}}\sim \frac{1}{M_{KK}}\;.
\end{equation}
On the other hand, the size of the holographic baryon
is
\begin{equation}
\rho_{baryon}\sim \frac{1}{M_{KK}\sqrt{\lambda }} \;.
\end{equation}
The difference is substantial in the large 't Hooft coupling limit
where this holographic QCD makes sense. Why is this?

Simply put, the Skyrmion solution of size $\sim 1/M_{KK}$ is a bad
approximation, because it solves the chiral Lagrangian which neglects
all other spin one mesons. This truncation can be justified for
processes involving low energy pions. The baryon is, however, a
heavy object and contains highly excited modes of pions,
and will excite relatively light vector mesons as well since
$U$ is coupled to vector and axial-vector mesons nontrivially at cubic
level. Therefore, the truncation to the pion sector is not a good
approximation as far as solitonic baryons are concerned, especially for
large 't Hooft coupling constant.\footnote{There
were previous studies that incorporated the effect of coupling
a single vector meson, namely the lightest $\rho$ meson, on the Skyrmion
which showed a slight shrinkage of the soliton \cite{vector-skyrmion}
as we would have expected in retrospect.}
We emphasize this difference because many of existing computation of
the baryon physics based on the Skyrmion picture must be thus rethought
in terms of the new instanton soliton picture. We will consider implication
of this new picture of the baryon in next sections.

Our solitonic picture of the baryon has a close tie to the usual AdS/CFT picture of
baryons as wrapped D-branes.
A  D4 brane wrapped along the compact $S^4$  corresponds to a
baryon vertex on the five-dimensional spacetime \cite{sakai-sugimoto},
as follows from an argument originally
due to Witten \cite{witten:1998xy}. To distinguish them from
the D4 branes supporting QCD, let us call them  D4$'$. On the D4$'$ worldvolume
we have again a Chern-Simons coupling of the form,
\begin{equation}
\mu_4\int C_3\wedge 2\pi\alpha'{d{\cal A}'}
\end{equation}
with D4$'$ gauge field ${ \cal A}'$, which can be evaluated
over $S^4$ as
\begin{equation}
2\pi\alpha'\mu_4\int_{S^4\times R}dC_3\wedge {\cal A}'=N_c\int_R {\cal A}' \;,
\end{equation}
where $R$ denotes the worldline in the noncompact part of the spacetime.
This shows that the
background $dC_3$ flux over $S^4$ induces $N_c$ unit of the electric
charge. On the other hand, the
Gauss constraint for ${\cal A}'$ demands that the net charge
should be zero, so the wrapped D4$'$ can exist only if $N_c$ end points
of fundamental strings are attached to D4$'$
to cancel this charge. In turn, the other ends of the
fundamental strings must go somewhere, and the only place it can
go is D8 branes. One can think of these strings as individual quarks that
constitute the baryon. Also, because of these fundamental strings,
the wrapped D4$'$ cannot be separated from D8's without a lot of
energy cost. The lowest energy state would be one where D4$'$ is
on top of D8's, which then would smear out as an instanton.
The latter is exactly the instanton soliton of ours.

\subsection{Quantum Numbers}

For the sake of simplicity, and also because
the quarks in this model have no bare mass,
we will take $N_F=2$ for the rest of the note.
A unit instanton soliton in question comes with six
collective coordinates. Three correspond to the position in $R^3$, and
three correspond to the gauge angles in $SU(N_F=2)$. If the soliton is
small enough ($\rho M_{KK}\ll 1$), there exists
approximate symmetries $SO(4)=SU(2)_+\times SU(2)_-$ at $w=0$, so the total rotational symmetry
of a small solution at origin is $SU(N_F=2)\times SU(2)_+\times SU(2)_-$.
Let us first see how the quantized instanton soliton fit into
representations of this approximate symmetry group.

The instanton can be rotated by an conjugate $SU(2)$ action as,
\begin{equation}
F\quad\rightarrow\quad S^\dagger FS \:,
\end{equation}
with any $2\times 2$ special unitary matrices $S$ which span
${\mathbf S}^3$.\footnote{Since $S$ and $-S$
rotates the solution the same way the moduli space is naively
${\mathbf S}^3/Z_2$. However at quantum level, we must consider states
odd under this $Z_2$ as well, so the moduli space is ${\mathbf S}^3$.}
Then, the quantization of the soliton is a matter of finding eigenstates
of free and nonrelativistic nonlinear sigma-model onto ${\mathbf S}^3$
\cite{Finkelstein:1968hy}.
$S$ itself admits an $SO(4)$ symmetry of its own,
\begin{equation}
S\quad\rightarrow \quad U S V^\dagger \;.
\end{equation}
Because of the way the spatial indices are locked with the
gauge indices, these two rotations are each identified as
the gauge rotation, $SU(N_F=2)$, and half of the spatial rotations, say,
$ SU(2)_+$.
Eigenstates on ${\mathbf S}^3$ are then nothing but the familiar
angular momentum eigenfunctions of three Euler angles,
conventionally denoted as
\begin{equation}
\vert s:p,q\rangle \; .
\end{equation}
Recall that the quadratic Casimirs of the two $SU(2)$'s (associated
with $U$ and $V$ rotations) always coincide to be $s(s+1)$.
One can proceed exactly in the same manner for anti-instantons, where
$SU(2)_+$ is replaced by $SU(2)_-$.

Therefore, under $SU(N_F=2)\times SU(2)_+\times SU(2)_-$,
the quantized instantons are in \cite{Park:2008sp}
\begin{equation}
(2s+1;2s+1;1) \;,
\end{equation}
while the quantized anti-instantons are in
\begin{equation}
(2s+1;1;2s+1) \;.
\end{equation}
Possible values for $s$ are integers and half-integers. However,
we are eventually interested in $N_c=3$, in which case spins and
isospins are naturally half-integral.  Thus we will subsequently
consider the case of fermionic states only. Exciting these isospin come at energy cost.
See Hata et.al. \cite{Hata:2007mb} for  mass spectra of some
excited instanton solitons.

\section{Holographic Dynamics}

The solitonic baryon is a coherent object which is made up of
pions as well as of vector and axial-vector mesons. This implies that
the structure of the soliton itself contains all the information on
how the baryon interacts with these infinite tower of mesons. This sort
of approach has been also used \cite{ANW} in the Skyrmion picture of old days, where,
for instance, the leading axial coupling for a nucleon emitting a soft
pion was computed following such thoughts. The difference here is that, instead of just
pions, all spin one mesons enter  this holographic construction
of the baryon, and this enables us to compute all low energy meson-hadron
vertices simultaneously.

\subsection{Dynamics of Hairy Solitons: Generalities}

First, we would like to illustrate the point by considering another
kind of solitons. The magnetic monopoles \cite{'tHooft:1974qc} appear as solitons in
non-Abelian Yang-Mills theories spontaneously broken to a subgroup
containing a $U(1)$ factor, such as in $SU(2)\rightarrow U(1)$,
and carries a magnetic charge. Usually it is a big and fluffy object
and must be treated as a classical object. However, if we push the
electric Yang-Mills coupling to be large enough, so that the magnetic
monopole size become comparable or even smaller than the symmetry breaking
scale, we have no choice but to treat it as a point-like object. The effective
action for this monopole field ${\cal M}$ (spinless for example) should
contain at least,
\begin{equation}
\biggl|\left(\partial_\mu +i\frac{4\pi}{e}\tilde A_\mu\right){\cal M}\biggr|^2 \;,
\end{equation}
where $\tilde A$ is the dual photon of the unbroken $U(1)$ gauge field. We know
this coupling exists simply because the  monopole
has the magnetic charge $4\pi/e$. But how do we know the latter fact?
Because the soliton solution itself exhibits  a long range
magnetic Coulomb tail of the form
\begin{equation}
F^{monopole}\sim\frac{4\pi}{e}\frac{1}{r^2} \;.
\end{equation}
If we replace the solitonic monopole by the quanta of the field
${\cal M}$ but do not couple to the dual photon field as above,
we would end up with a local excitation. However, a magnetic monopole
(or an electrically charge particle) is not really a local object. Creating one
always induces the corresponding long range magnetic (electric) Coulomb field.
To ensure that the effective field theory represent the magnetic monopole
accurately, we must make sure that creating a quanta of ${\cal M}$ is
always followed by creation of the necessary magnetic Coulomb field.
This is achieved by coupling the local field ${\cal M}$ to the gauge field
$\tilde A$ at an appropriate strength.
This is a somewhat unconventional way to understand
the origin of the minimal coupling of the monopole to the dual
gauge field $\tilde A$.

\subsection{The Small Size Matters}

Before going further, let us
briefly pose and ask about the validity of such an approach for our
solitonic baryon. The
key to this is a set of inequalities among three natural scales that
enter the baryon physics, which are
\begin{equation}
\frac{1}{M_{KK}}\gg\frac{1}{M_{KK}\sqrt{\lambda}}\gg\frac{1}{M_{KK}N_c\lambda}\;.
\end{equation}
They hold in the large $N_c$ and large $\lambda$ limit.
The first is the length scale of mesons, the second is the classical
size of the solitonic baryon, and the third is the Compton
wavelength of the baryon since its mass is $\sim M_{KK}N_c\lambda$.

The first inequality tells us that the baryon tends to be much smaller
than mesons and thus can be regarded almost pointlike when interacting
with mesons. This justifies the effective field theory approach
where we think of each baryon as small excitation of a field. One does
this precisely when the object in question can be treated as if it has no internal
structure other than quantum numbers like spins.

The second inequality tells us that the quantum uncertainty associated
with the baryon is far smaller than the classical core size of the soliton.
This is important because, otherwise, one may not be able to trust anything
about the classical features of the soliton at quantum level. When
the second inequality holds, it enables us to make use of
the classical shape of the soliton and to extract information about how
meson interact with the baryon. The fact we have a small soliton size and
an even smaller Compton size of that soliton is very fortunate.

\subsection{Holographic Dynamics of Baryons}

As with the small magnetic monopole case, we wish to trade off
the (quantized) instanton soliton in favor of local baryon field(s) and make sure
to encode the long-range  tails of the soliton in how
the baryon field(s) interacts with the low energy gauge fields.
Our instanton soliton has two types of distinct but related long-range
field. The first is due to the Pontryagin density and goes like
\begin{equation}\label{long}
F_{mn}\sim \frac{\rho^2_{baryon}}{(r^2+w^2)^2} \;,
\end{equation}
while the second is the Coulomb field due to the Chern-Simons coupling
between ${\cal A}$ and $F\wedge F$,
\begin{equation}
{\cal F}_{0n}\sim \frac{e(w)^2N_c}{(r^2+w^2)^{3/2}}\;.
\end{equation}
The latter is the five-dimensional analog of the electric Coulomb tail.

Apart from the fact that we have two kinds of long-range fields,
there is another important difference from the monopole case.
AS we saw in section 3.2, the solitonic baryon has ${\mathbf S}^3$
worth of internal moduli, quantization of which gave us the various
spin/isospin baryons. Since the gauge direction of the magnetic
long range field is determined by coordinate on ${\mathbf S}^3$,
the field strengths associated with the Pontryagin density
should be smeared out by quantum fluctuation along the moduli space.
It is crucial for our purpose that what we mean by
long-range fields of the instanton soliton are actually
these quantum counterpart, not the naive classical one.
Basic features of the smearing out effect and relevant
identities can be found in next subsection.

The electric Coulomb tail should be encoded in a minimal coupling
to the Abelian part of ${\cal A}$. For a spin/isospin half Baryon,
${\cal B}$, we anticipate a minimal term of the form
\begin{equation}
\bar{\cal B}
(N_c {\cal A}_m^{U(1)}+{A}_\mu)\gamma^m{\cal B} \;.
\end{equation}
This is uniquely fixed by the Coulomb charge $N_c$
and the $SU(N_F=2)$ representation of the quantized instanton.
The purely magnetic tail of the soliton is more subtle to deal with.
From the simple power counting, it is obvious that the coupling
responsible for such a tail must have one higher dimension than
the minimal coupling, hinting at the field strength $F$  of the
$SU(N_F=2)$ part coupling directly to a baryon bilinear, such as
\begin{equation}
\bar{\cal B}{F}_{mn}\gamma^{mn}{\cal B} \;.
\end{equation}
It turns out that this is precisely the right structure to
mimic the long-range magnetic fields of quantized instantons
and anti-instantons.\footnote{
In fact, a prototype of this simple method makes a brief appearance in
the landmark work on Skyrmion by Adkins, Nappi, and Witten \cite{ANW}. In their case,
however, this gives only the pion-baryon interactions, forcing them to
a related but somewhat different formulation. In our case, this method
generates all meson-baryon interactions, however.}

To show that the latter vertex is indeed the precisely
right one, one must consider the following  points.
(1) Is this the unique term that can reproduce the correct
quantum-smeared long-range instanton and anti-instanton tail?
(2) If so, how do we fix the coefficient, taking into account the quantum
effects. (3) And is the estimate reliable? The answers are long and technical.
We refer the readers to  literatures \cite{Hong:2007kx,Hong:2007ay,Park:2008sp}
for precise answers to these questions,
but here state that the answers are all affirmative and that
the effective action of mesons and baryons is uniquely determined by
this simple consideration. This is true at least in the large $N_c$
and the large $\lambda$ limit.

This leads to the following five-dimensional
effective action,
\begin{eqnarray}
&&\int d^4 x dw\left[-i\bar{\cal B}\gamma^m D_m {\cal B}
-i m_{\cal B}(w)\bar{\cal B}{\cal B} +{2\pi^2\rho_{baryon}^2\over
3e^2(w)}\bar{\cal B}\gamma^{mn}F_{mn}{\cal B} \right]\nonumber \\
&-&\int d^4 x  dw {1\over 4 e^2(w)} \;{\rm tr}\, {\cal F}_{mn}{\cal F}^{mn}\,,
\label{5d}
\end{eqnarray}
with the covariant derivative given as $D_m=\partial_m-i(N_c {\cal A}_m^{U(1)}+{A}_m)$
with $A_m$ in the fundamental representation of $SU(N_F=2)$.

The position-dependent mass $m_{\cal B}(w)\sim 1/e(w)^2$ is a very sharp
increasing function of $|w|$, such that in the large $N_c$ and large $\lambda$ limit,
the baryons wavefunction is effectively localized at $w=0$. This is the limit
where the above effective action is trustworthy. We find
\begin{equation}
{2\pi^2\rho_{baryon}^2\over 3e^2(0)} = \frac{N_c}{\sqrt{30}}\cdot\frac{1}{M_{KK}}\;,
\end{equation}
so the last term involving baryons can be actually dominant over the minimal coupling,
despite that it looks subleading in the derivative expansion. As it turns out,
this term is dominant for cubic vertex processes involving pions or axial
vector mesons, whereas the minimal coupling dominates for those involving vector mesons
\cite{Hong:2007ay}.

\subsection{Basic Identities and Isospin-Dependence}

We have discussed general ideas behind the effective action approach
and given the explicit results for isospin 1/2 case. The only term that is
not obvious is the coupling between baryons and the field strength $F$,
with the coefficient
${2\pi^2\rho_{baryon}^2/3e^2(0)}$, and we would like to spend a little more
time on its origin. Apart from convincing readers that the derivation of
the effective action is actually rigorous, this would also allow us to outline
how the result generalizes for higher isospin baryons, such $\Delta$ particles,
as well.

Each and every quantum of the baryon field ${\cal B}$ is supposed
to represent a quantized (anti-)instanton soliton. Let us recall that
the quantization of the soliton involves finding wavefunctions on the
moduli space of the soliton, which is $S^3$. Since the moduli encode
the gauge direction of the instanton soliton, the classical gauge
field is quantum mechanically smeared and should be replaced by
its expectation values as
\begin{equation}\label{smeared}
F\;\rightarrow\;\langle\langle S^\dagger FS\rangle\rangle =
\langle\langle\Sigma_{ab} \rangle\rangle F^b\;,
\end{equation}
with $2\Sigma_{ab} \equiv {\rm tr}\left[\tau_a S^\dagger
{\tau_b}S \right]$. $\langle\langle\cdots\rangle\rangle$ means taking expectation
value on wavefunctions on the moduli space of the soliton, and computes the
quantum smearing effect.

The effective action (\ref{5d}) would make sense if and only if
each quanta of the baryon field ${\cal B}$ is equipped with precisely
the right smeared-out gauge field of this type. How is this possible?
For the simplest case of isospin 1/2,  the relevant identity that
shows this reads\footnote{
This identity for $s=1/2$ is originally due to Adkins, Nappi, and Witten,
who obtained it in the context of the Skyrmion. The moduli space of a Skyrmion
and that of our instanton soliton coincides, so the same identity holds.}
\begin{equation}\label{anw}
\langle\langle 1/2:p',q'\vert\,\Sigma_{ab}\vert 1/2:p,q\rangle\rangle
=-\frac{1}{3}({\cal U}(1/2:p',q')^{\epsilon'}_{\beta'})^*
  \sigma_a^{\beta' \beta} \tau_b^{\epsilon'\epsilon}
{\cal U}(1/2:p,q)^{\epsilon}_{\beta}
\end{equation}
where ${\cal U}(1/2:p,q) $ is the two-component spinor/isospinor
of $J_3=p$, $I_3=q$, and $J^2=I^2=3/4$. Identifying the two-component
spinor ${\cal U}$ as the upper half of the four-component spinor ${\cal B}$
representing positive energy states, one can show that the equation of
motion for the gauge field coupled to ${\cal B}$ is
\begin{equation}\label{ymeom}
(\nabla\cdot F)_m^a\sim \nabla_n\left(\bar\eta^b_{nm}
{\cal U}^\dagger(\sigma_b\tau^a){\cal U}\right)+\cdots
\;,
\end{equation}
which shows, via (\ref{anw}), that the quanta ${\cal U}$ of
${\cal B}$ would be accompanied by the correctly smeared long range tail
of gauge field of type (\ref{smeared}). The right hand side
comes from the coupling of type
\begin{equation}
\bar {\cal B}F{\cal B}
\end{equation}
in (\ref{5d}). A similar match can be shown
for negative energy states, where the 't Hooft symbol $\bar\eta$ is replaced
by $\eta$ and ${\cal U}$ by its anti-particle counterpart ${\cal V}$.  A careful check
of the normalization leads us to the coefficient ${2\pi^2\rho_{baryon}^2/ 3e^2(0)}$,
where the number $3$ in the denominator came from the factor $1/3$ in
Eq.~(\ref{anw}).

It turns out that this goes beyond $s=1/2$. The
identity (\ref{anw}) is generalized to
for arbitrary half-integral $s$ as \cite{Park:2008sp}
\begin{eqnarray}\label{anw2}
\langle\langle s:p',q'\vert\Sigma_{ab}\vert s:p,q\rangle\rangle
=-\frac{s}{s+1}\cdot({\cal U}(s:p',q')^{\epsilon'\epsilon_2 \cdots
  \epsilon_{2s}}_{\beta'\alpha_2 \cdots \alpha_{2s}})^*
  \sigma_a^{\beta' \beta} \tau_b^{\epsilon'\epsilon}
{\cal U}(s:p,q)^{\epsilon\epsilon_2 \cdots
  \epsilon_{2s}}_{\beta \alpha_2 \cdots \alpha_{2s}}\;,
\end{eqnarray}
where the left-hand-side is again evaluated as wavefunction-overlap integral
on the moduli space $S^3$ of the instanton soliton. ${\cal U}$ is now that
of higher spin/isospin field with symmetrized multi-spinor/multi-isospinor indices.
As with ${\cal U}(1/2)$,  ${\cal U}(s)$'s are positive energy spinors with
each index taking values 1 and 2. This implies a cubic interaction term
of type
\begin{equation}
\bar{\cal B}_sF{\cal B}_s
\end{equation}
where ${\cal B}_s $ denotes a local baryon field of isospin $s$ and
$SO(4)=SU(2)_+\times SU(2)_-$ angular momentum $[s]_+\otimes[0]_-\oplus [0]_+\otimes[s]_-$.
Relative to the isospin 1/2 case, the coefficient is increased
from $1/3$ to $s/(s+1)$, which reflects the obvious fact that higher
angular momentum states would be less and less smeared.

Finally, with $s>1/2$ baryons included, there are one more type of
processes allowed where a baryon changes its own isospin by emitting
isospin 1 mesons. The relevant identities for these processes are
\begin{equation}\label{ss+1}
\langle\langle s:p',q'\vert\Sigma_{ab}\vert s+1:p,q\rangle\rangle
=-\frac12\sqrt{\frac{2s+1}{2s+3}}\cdot\left[{{\cal U}(s:p',q')^\dagger}{\cal U}(s+1:p,q)_{ab}\right] \;,
\end{equation}
where $3\times 3$  spin/isospin $s$ wavefunctions ${\cal U}(s+1:p,q)_{ab}$ are
\begin{eqnarray}
\left({\cal U}(s+1:p,q)_{ab}\right)_{\alpha_1\cdots \alpha_{2s}}^{\epsilon_1\cdots \epsilon_{2s}}
\equiv
(\sigma_2\sigma_a)^{\beta\beta'}(\tau_2\tau_b)_{\epsilon\epsilon'}
{\cal U}(s+1:p,q)_{\beta\beta' \alpha_1\cdots \alpha_{2s}}^{\epsilon\epsilon'
\epsilon_1\cdots \epsilon_{2s}}\;.
\end{eqnarray}
This shows up in the effective action of baryon as a coupling of
type
\begin{equation}
\bar{\cal B}_{s+1}F{\cal B}_s\;.
\end{equation}
The complete effective action of baryons with such arbitrary half-integer
isospins was given in
Ref.~\cite{Park:2008sp}. For the rest of the note, we will confine ourselves to
isospin 1/2 case.

\section{Nucleons}

Nucleons are the lowest lying baryons with isospin and spin 1/2. As such,
they arise from the isospin 1/2 holographic baryon field ${\cal B}$ whose
effective action is given explicitly above. This effective action
contains interaction terms between currents of ${\cal B}$ with the $U(N_F)$
gauge field of five dimensions, and thus contain an infinite number of
interaction terms between nucleons and mesons, specifically all cubic
couplings involving nucleons emitting pion, vector mesons, or axial-vector
mesons. Extracting four-dimensional amplitudes of interests is a simple matter of
dimensional reduction from $ R^{3+1}\times I $ to $R^{3+1}$. In this
section, we show this procedure, showcase some of the simplest examples
such comparisons, and comment on how the results should be taken in view
of various approximation schemes we relied on.

\subsection{Nucleon-Meson Effective Actions}

The effective action for the four-dimensional nucleons is derived from
this, by identifying the lowest eigenmode of ${\cal B}$ upon the
KK reduction along $w$ direction as the
proton and the neutron. Higher KK modes would be also isospin half
baryons, but the gap between the ground state and excited state is
very large in the holographic limit, so we consider only the
ground state. We mode expand ${\cal B}_\pm(x^\mu,w)={\cal
N}_\pm (x^\mu)f_\pm (w)$, where $\pm$ refers to the chirality
along $w$ direction, and reconsitute a four-dimensional
spinor ${\cal N}$ with $\gamma^5 {\cal N}_\pm=\pm {\cal N}_\pm$ as its
chiral and anti-chiral components. The lowest KK eigenmodes $f_\pm(w)$
solve
\begin{eqnarray}
&&\left[-\partial^2_w \mp \partial_w m_{\cal B}(w)+(m_{\cal B}(w))^2\right]
f_\pm(w)=m_{\cal N}^2 f_\pm(w)\:,\label{eigeneq}
\end{eqnarray}
with some minimum eigenvalue $m_{\cal N}> m_{\cal B}(0)=m_{\cal B}^{classical}$.
This nucleon mass $m_{\cal N}$ will generally differ from the five-dimensional
soliton mass $m_{\cal B}^{classical}$, due to quantization of light modes
such as spread of the wavefunction $f_{L,R}$ along the fifth direction.

Inserting this  into the action (\ref{5d}),  we find the following structure
of the four-dimensional nucleon action
\begin{equation}
\int dx^4\;{\cal L}_4 = \int dx^4\left(-i\bar {\cal N}
\gamma^\mu\partial_\mu {\cal N}-im_{\cal N}\bar {\cal N}{\cal N}+ {\cal
L}_{\rm vector} +{\cal L}_{\rm axial}\right)\:,
\end{equation}
where we have, schematically, the vector-like couplings
\begin{equation}
{\cal L}_{\rm vector}=-i\bar {\cal N} \gamma^\mu \beta_\mu
 {\cal N}-\sum_{k\ge 0}g_{V}^{(k)} \bar {\cal N} \gamma^\mu  a_\mu^{(2k+1)}
 {\cal N}\:,\label{vector-coupling}
\end{equation}
and the axial couplings to axial mesons,
\begin{equation}
{\cal L}_{\rm axial}=-\frac{i g_A}{2}\bar {\cal N}  \gamma^\mu\gamma^5
\alpha_\mu {\cal N} -\sum_{k\ge 1} g_A^{(k)} \bar {\cal N} \gamma^\mu\gamma^5
a_\mu^{(2k)} {\cal N}\:.
\end{equation}
All the  couplings constants $g_{V,A}^{(k)}$ and $g_A$ are calculated by
suitable wave-function overlap integrals involving $f_\pm$ and $\psi_{(n)}$'s.

Although we did not write so explicitly, isospin triplet mesons and singlet
mesons have different coupling strength to the nucleons, so there are actually
two sets of couplings $(g_A, g_A^{(k)},g_{V}^{(k)})$, one for isosinglet
mesons, such as $\omega$ and $\eta'$, and the other for isotriplet mesons,
such as $\rho$ and $\pi$.
The leading contribution to axial couplings in the isospin triplet channel
arise from the direct coupling to $F_{mn}$, and are all proportional to
$\rho^2_{baryon}$. All the rest are dominated by terms from the five-dimensional
minimal coupling to ${\cal A}_m$. We refer interested readers to
Ref.~\cite{Hong:2007kx,Hong:2007ay} for explicit form of these coupling
constants.

\subsection{Numbers and Comments}

To showcase typical predictions from the above setup, let us quote
two notable examples for the nucleons \cite{Hong:2007ay}. The first is the cubic coupling
of the  lightest vector mesons to the nucleon, to be denoted as
$g_{\rho{\cal NN}}$ for the isotriplet meson $\rho$ and $g_{\omega\cal NN}$
for the iso-singlet meson $\omega$. In the above effective action, these two
are denoted collectively as $g_V^{(0)}$. An interesting prediction of
this holographic effective action of nucleons is that
\begin{equation}
\frac{g_{\omega\cal NN}}{g_{\rho\cal NN}}=N_c+\delta \;,
\end{equation}
where the leading $N_c$ is a consequence from the five-dimensional
minimal coupling to ${\cal A}$ while the subleading correction $\delta$
arises from the direct coupling to the field strength $F$. With
$N_c=3$ and $\lambda\simeq 17$ (the latter is required by fitting $f_\pi$
and $M_{KK}$ to the pion decay constant and the vector meson masses
to actual QCD), we find
\begin{equation}
\frac{g_{\omega\cal NN}}{g_{\rho\cal NN}}\simeq 3+0.6 =3.6 \;.
\end{equation}
Extracting ratios like this from experimental data is somewhat
model-dependent, with no obvious consensus, but the ratio is
believed to be larger than 3 and numbers around 4-5 are typically found.
Given the crude nature of our approximation and that there is
no tunable parameter other than the QCD scale and $f_\pi$,
the agreement is uncanny. A more complete list of various cubic
couplings between spin one mesons and nucleons has been worked
out in Ref.~\cite{Hong:2007ay} and further elaborated recently
in Ref.~\cite{Deuteron}.

The leading axial coupling to pions, $g_A$, is somewhat better
measured at $\simeq 1.26$. Our prediction is \cite{Hong:2007kx}
\begin{equation}
g_A=\frac{2\lambda
N_c(\rho_{baryon}M_{KK})^2}{ 81\pi^2} +\cdots =
\left(\frac{24}{5\pi^2}\right)^{1/2}\times\frac{N_c}{3}+\cdots \;,
\end{equation}
where the leading term arise from the direct coupling to the field strength $F$
and the ellipsis denotes the subleading and higher correction. While this
does not look too good, we must remember that this holographic model
is effectively a quenched QCD, missing out on possible $O(1)$
corrections. From old studies of large $N_c$ constituent models,
a group theoretical $O(1)$ correction has been proposed for this
type of operators, which states that the next leading correction
would amounts to $N_c\rightarrow N_c+2$\footnote{See Ref.~\cite{Hong:2007ay}
for more explanations and references}. So, in a more realistic
version where we take into account of the backreaction of D8 branes
on the dual geometry, we may anticipate for $N_c=3$
\begin{equation}
g_A \simeq \left(\frac{24}{5\pi^2}\right)^{1/2}\times\frac{N_c+2}{3}+O(1/N_c)\simeq 1.16+O(1/N_c) \;.
\end{equation}
Finally, $O(1/N_c)$ is partly captured by the minimal coupling term
our quenched model, which turns out to give roughly
a 10\% positive correction, making the total very close
to the measured quantity 1.26.

These two illustrate nicely what kind of predictions can be made
and how accurate their predictions can be when compared to experimental
data. Much more rich array of predictions exist, such as other cubic
couplings between mesons and baryons,
anomalous magnetic moment \cite{Hong:2007kx}, complete vector dominance of electromagnetic
form factors \cite{Hong:2007ay}, and detailed prediction on momentum dependence of such
form factors \cite{Hong:2007dq,Hashimoto:2008zw}.

However,
one should be a bit more cautious. The model, as an approximation to
real QCD, has many potential defects. The main problem is that all of this
is in the context of large $N_c$ and that any prediction, such as above two,
has to involve an extensive extrapolation procedure. Many ambiguities
can be found in such a procedure, and we chose a particular strategy
of computing all quantities and analytically continuing the final
expressions for the amplitudes to realistic QCD regime. The fact
it works remarkably well does not really support its validity in any
rigorous sense. Also the D4-D8 model we employed include many massive
fields which are not part of ordinary four-dimensional QCD, and one
should be cautious in using the holographic QCD for physics other than
simple low energy processes.

Despite such worries, the D4-D8 holographic QCD turned out to be far
better than one may have anticipated. We have shown how it accommodates not only
the (vector) meson sector but the baryon sector very competently.\footnote{
 One of the acutely
missing story is how the spinless mesons (except Goldstone bosons) would fit in
the story. Initial investigation of this gave a possibly disappointing
result, although it may have more to do with how the lightest scalar
mesons are rather complicated objects and may not be a bi-quark meson
of conventional kind \cite{Sugimoto}.}
Whether or not the holographic QCD can be elevated to a controlled
and justifiable approximation to real QCD remains to be seen, depending
crucially on having a better understanding of the string theory in the curved
spacetime. Nevertheless, it is fair to say that we finally have a rough
grasp of the physics that controls {\it the master fields}, and perhaps this
insight by itself will lead to a better and more practical formulation
of the QCD in the future.

\section{Electromagnetic Properties}

Holographic baryons and their effective action in the bulk also encodes
how baryons, and in particular, nucleons would interact with electromagnetism.
For this, one follows the usual procedure of AdS/CFT where operators in the
field theory are matched up with non-normalizable modes of bulk fields.
Operationally, one simply introduces the boundary photon field ${\cal V}$
as a nonnormalizable mode, which adds to $\beta$-term in the expansion of ${\cal A}$,
\begin{equation}
A_\mu(x;w)= i\alpha_\mu(x)\psi_0(w)+{\cal V}_\mu(x)+i\beta_\mu(x) +\sum_n
a_\mu^{(n)}(x)\psi_{(n)}(w) \:,
\end{equation}
and repeat the dimensional reduction to the four dimensions. For instance,
computation of anomalous magnetic moments of proton and neutron can be done
with relative ease, and  gives remarkably good agreement with measured values
\cite{Hong:2007ay}.

For more detailed accounts of electromagnetic properties of baryon,
we refer the readers to Refs.~\cite{Hong:2007ay,Hong:2007dq,Hashimoto:2008zw}.
Here we will only consider the most notable feature of the electromagnetic
properties, namely the complete vector dominance, whereby all electromagnetic
interactions are entirely mediated by the infinite tower of vector mesons.
This also illustrate well how the holographic QCD can give a sweeping
and qualitative prediction and also where it could fail.

The vector dominance means that there is no point-like charge, which,
in view of the minimal coupling between ${\cal A}$ and ${\cal B}$ in
(\ref{5d}), sounds pretty odd. To understand what's going on, one must
consider quadratic structures in the vector meson sector. Defining
\begin{equation}
\zeta_k=\int dw\frac{1}{2e(w)^2}\,\psi_{(2k+1)}(w) \:,
\end{equation}
for parity even eigenfunctions $\psi_{(2k+1)}$'s,
the quadratic part of the vector meson is \cite{sakai-sugimoto}
\begin{equation}
\sum_k {\rm tr}\;\left[ -\frac12 |\,dv^{(k)}|^2 -
m_{(2k+1)}^2|\,v^{(k)}-\zeta_k ({\cal
V}+i\beta)|^2\right]\;,
\end{equation}
where we introduced the shifted vector fields
\begin{equation}
v^{(k)}=a^{(2k+1)}+\zeta_k ({\cal V}+i\beta) \:.
\end{equation}
This mixing of vector mesons and photon is at the heart
of the vector dominance.
(The axial-vector mesons, $a^{(2k)}$'s, do not mix with photon because
of the parity.)

Now let us see how this mixing of vector fields enters the coupling
of baryons with electromagnetic vector field ${\cal V}$. Taking the
minimal coupling, we find
\begin{equation}
\int dw \;\bar {\cal B}\gamma^mA_m{\cal B}=
\bar B\gamma^\mu {\cal V}_\mu B+\sum_k g^{(k)}_{V,min} \bar B\gamma^\mu a^{(2k+1)}_\mu B+\cdots \:,
\end{equation}
where the ellipsis denotes axial couplings to axial vectors
as well as  coupling to pions via $\alpha_\mu$ and $\beta_\mu$.
$g^{(k)}_{V,min} $ is the cubic coupling between $k$-th vector
meson and the baryon, or more precisely its leading contribution
coming from the minimal coupling to ${\cal A}$.
Again, the presence of the direct minimal coupling to the photon
${\cal V}$ seems to contradict the notion of vector dominance.
However, it is advantageous to employ the canonically normalized
vector fields $v^{(k)}$ in place of $a^{(k)}$, upon which this
becomes
\begin{equation}
\bar B\gamma^\mu {\cal V}_\mu B+\sum_k g^{(k)}_{V,min}
\bar B\gamma^\mu (v^{(k)}_\mu-\zeta_k {\cal V}_\mu ) B+\cdots \:.
\end{equation}
On the other hand,
\begin{eqnarray}
\sum_kg^{(k)}_{V,min}\zeta_k&=&\sum_k \int dw'\;|f_+(w')|^2\psi_{(2k+1)}(w')\times
\int dw\;\frac{1}{2e(w)^2}\,\psi_{(2k+1)}(w)\nonumber\\
&=& \int dw'\;|f_+(w')|^2\times
\int dw\;\delta(w-w') \;=\;1 \:,
\end{eqnarray}
where we made use of the definite parities of $1/e(w)^2$ and $\psi_{(n)}$'s
and also of the completeness of $\psi_{(n)}$'s. This sum rule
$\sum_kg^{(k)}_{V,min}\zeta_k =1$ forces
\begin{equation}
\bar B\gamma^\mu {\cal V}_\mu B+\sum_k g^{(k)}_{V,min} \bar B\gamma^\mu (v^{(k)}_\mu
-\zeta_k{\cal V}_\mu) B+\cdots
= \sum_k g^{(k)}_{V,min} \bar B\gamma^\mu v^{(k)}_\mu B+\cdots
\end{equation}
and the baryon couples to the photon field ${\cal V}$ only via
$v^{(k)}$'s which mixes with ${\cal V}$ in their mass terms.

This choice of basis is only for the sake of clarity. Regardless of
the basis, the above shows that no coupling between ${\cal V}$ and
${\cal B}$ can occur in the infinite momentum limit. This statement
is clear in the $\{{\cal V}; v^{(k)}\}$ basis which is diagonal
if the mass term is negligible. Alternatively,
we can ask for the invariant amplitude of the charge form factor,
to which the minimal coupling contributes \cite{Hong:2007ay}
\begin{equation}
F_{1,min}(q^2)=1-\sum_k \frac{g^{(k)}_{V,min}\zeta_k
q^2}{q^2+m_{(2k+1)}^2}= \sum_k \frac{g^{(k)}_{V,min}\zeta_k m_{(2k+1)}^2
}{q^2+m_{(2k+1)}^2} \label{chargeff}
\end{equation}
with the momentum transfer $q$. For small momentum transfer, the
first few light vector mesons dominate the form factors by mediating
betwee the baryon and the photon. This end fit
with experimental data pretty well.
Similar computation can be done for the magnetic form factor, from
which one also finds the (anomalous) magnetic moment that fits the
data pretty well \cite{Hong:2007ay,Hong:2007kx}.

However, for large momentum transfer, the form factor decays as $1/q^2$
which is actually too slow for real QCD baryons. Estimates based on the
parton picture say that the decay should be $\sim 1/q^{2(N_c-1)}$.
This dramatic failure of the form factor for large momentum regime should
not be a big surprise. The theory we started with is a low energy limit
of D4-D8 complex compactified  (with warp factors) on $S^1\times S^4$.
As such, one has to truncate infinite number of massive modes
in order to reach a QCD-like theory in the boundary and must stay
away from that cut-off scale to be safe from this procedure. For
large momentum transfers, say larger than $M_{KK}$, the computation
we relied on has no real rationale. This should caution readers that
the holographic QCD, at least in the limited forms that are available now,
is not a fix for everything. One really must view it as a vastly
improved version of the chiral Lagrangian approach, with many hidden
symmetries now manifest, but still suitable only for low energy physics.

\section{More Comments}

D4-D8 holographic model of QCD is  the most successful
model of its kind known. It reproduces in particular detailed
particle physics of mesons and baryons. One reason for its success
can be found in the fact that it builds on the the meson sector,
the lightest of which is lighter than the natural cut-off scale $M_{KK}$.
Apart from $1/N_c$ and $1/\lambda$ expansions imposed by general
AdS/CFT ideas, one also must be careful with low energy expansion
as well, because, we we stated before, the model includes many
more massive Kaluza-Klein modes and even string modes that
are not part of ordinary QCD. For low energy processes, nevertheless,
one would hope that these extra massive states (above $M_{KK}$)
do not contribute too much, which seems to be the case for low
lying meson sector \cite{sakai-sugimoto}.

Our solitonic and holographic model of baryons elevates the
classic Skyrme picture based on pions to a unified model
involving all spin one mesons in addition to pions. This
is why the picture is extremely predictive.
As we saw in this note, for low momentum processes, such as soft
pion processes, soft rho meson exchanges, and soft elastic
scattering of photons, the model's predictions compare extremely well
with experimental data. It is somewhat mysterious that the baryon
sector works out almost as well as the meson sector, since baryons
are much heavier than  $M_{KK}$ in the large $N_c$ and the
large $\lambda$ limit.

Note that the soliton underlying the baryon is nearly
self-dual in the large $\lambda$ limit. For instance, Eq.~(\ref{mass})
shows that the leading, would-be BPS, mass is dominant over
the rest by a factor of $\lambda$. There must be a sense
in which the soliton is approximately supersymmetric with
respect to the underlying IIA string theory, even though the
background itself breaks all supersymmetry at scale $M_{KK}$.
One may argue that even though there are many KK modes
and even stringy modes lying between the naive cut-off scale $M_{KK}$
and the baryon mass scale $M_{KK}N_c\lambda$, these non-QCD degrees
of freedom would be paired into approximate supermultiplets, reducing
their potentially destructive effect, especially because the baryon
itself is roughly BPS. Whether or not one can actually quantify such
an idea for the model we have is unclear, but if possible it would be
an important step toward rigorously validating holographic approaches to
baryons in this D4-D8 set-up.

There are more work to be done. One important direction is to  perform
more refined
comparisons against experiments. In particular, extracting coupling
constants from raw data seems quite dependent on theoretical models,
and it is important to compute directly measurable amplitudes starting
from the effective action of ours. Nucleon-nucleon scattering amplitudes
or more importantly the nucleon-nucleon potential would be a good place to
start \cite{Kim:2009iy,Hashimoto:2009ys,Deuteron}.
Another profitable path would be to
consider dense system such as neutron stars as well as physics of
light nuclei, where our model with
far less tunable parameters would give unambiguous predictions.
This will in turn further test the model as well.

\section*{Acknowledgements}
This note is based on a set of collaborative works with D.K. Hong, J. Park, M. Rho, and H.-U. Yee.
The author wishes to thank SITP of Stanford University, Aspen Center for Physics, and also
organizers of the conference ``30 years of mathematical method in high energy physics"
for hospitality. This work is supported in part by the Science Research Center Program
of KOSEF (CQUeST, R11-2005-021), the Korea Research Foundation (KRF-2007-314-C00052),
and by the Stanford Institute for Theoretical Physics (SITP Quantum Gravity visitor fund).

\end{document}